\documentclass[floatfix,notitlepage,prd,superscriptaddress,twocolumn,nofootinbib]{revtex4-1}
\usepackage{graphicx,amsmath,bm,color,epsfig,natbib}
\usepackage[pdfencoding=auto]{hyperref}

\usepackage[T1]{fontenc} 
\newcommand{\dd}{\mathrm{d}}
\newcommand{\lsim}{\;\mbox{\raisebox{-0.5ex}{$\stackrel{<}{\scriptstyle{\sim}}$}
}\;}

\begin{document}

\title{Splashback in galaxy clusters as a probe of cosmic expansion and gravity}

\author{Susmita Adhikari}
\affiliation{Kavli Institute for Particle Astrophysics and Cosmology, Stanford University, 452 Lomita Mall Stanford, CA 94305, USA}
\email{susmita@stanford.edu}
\affiliation{Department of Astronomy, University of Illinois at Urbana-Champaign, 1002 W. Green Street, Urbana, IL 61801, USA}
\author{Jeremy Sakstein}
\affiliation{Department of Physics and Astronomy, University of Pennsylvania, 209 S. 33rd St. Philadelphia, PA 19104, USA}
\email{sakstein@physics.upenn.edu, bjain@physics.upenn.edu}
\author{Bhuvnesh Jain}
\affiliation{Department of Physics and Astronomy, University of Pennsylvania, 209 S. 33rd St. Philadelphia, PA 19104, USA}
\author{Neal Dalal}
\affiliation{Perimeter Institute for Theoretical Physics, 31 Caroline St N, Waterloo, ON N2L 2Y5, Canada }
\email{ndalal@perimeterinstitute.ca}
\author{Baojiu Li}
\affiliation{Institute for Computational Cosmology, Department of Physics, Durham University, Durham DH1 3LE, UK}
\email{baojiu.li@durham.ac.uk}

\begin{abstract}

The splashback radius is a physical scale in dark matter halos that is set by the gravitational dynamics of recently accreted shells. We use analytical models and N-body simulations to study the dependence of splashback on dark energy and screened modified gravity theories. In modified gravity models, the transition from screened to unscreened regions typically occurs in the cluster outskirts, suggesting potentially observable signatures in the splashback feature.  We investigate the location of splashback in both chameleon and Vainshtein screened models and find significant differences compared with $\Lambda$CDM predictions. We also find an interesting interplay between dynamical friction and modified gravity, providing a distinctive signature for modified gravity models in the behavior of the splashback feature as a function of galaxy luminosity. 
\end{abstract}

\maketitle
\flushbottom
\section{Introduction} \label{intro} 

The accelerated expansion of the universe requires either the introduction of a dark energy component to the energy density in Einstein's field equations or a resolution of the UV-sensitivity and radiative instability of the cosmological constant \cite{Kaloper:2015jra,Padilla:2015aaa,Khoury:2018vdv}. This component does not significantly cluster gravitationally \cite{Copeland:2006wr}, and if its equation of state parameter $w=P/\rho$ is consistent with $w=-1$, then dark energy is consistent with a cosmological constant, $\Lambda$. The standard $\Lambda$CDM cosmological model assumes that dark energy is equivalent to a cosmological constant, and this model has been highly successful in explaining a vast array of observable properties of the large-scale universe. An alternative to $\Lambda$, however, is to account for dark energy not by adding a vacuum term to the Einstein equations, but instead by modifying gravity from general relativity on large scales. While general relativity has been tested extensively on solar system scales \cite{Bertotti:2003rm,Brax:2013uh,Will:2014kxa,Ip:2015qsa,Burrage:2016bwy,Burrage:2017qrf,Sakstein:2017pqi}, astrophysical objects \cite{Davis:2011qf,Jain:2012tn,Sakstein:2013pda,Vikram:2014uza,Sakstein:2014nfa,Sakstein:2015oqa,Koyama:2015oma,Sakstein:2015zoa,Sakstein:2015aac,Sakstein:2015aqx,Sakstein:2017bws}, and in the strong-field regime \cite{Weisberg:2010zz,Babichev:2016jom,Sakstein:2016oel,Abbott:2016blz,TheLIGOScientific:2016src,GBM:2017lvd,Sakstein:2017xjx,Baker:2017hug,Creminelli:2017sry,Ezquiaga:2017ekz}, tests of general relativity (GR) on cosmological scales are not yet conclusive \cite{Alonso:2016suf}. By construction, modified gravity models for dark energy can produce the same large-scale expansion history as scalar field models like quintessence or $\Lambda$, which means that observationally distinguishing between these theories requires searching for signatures on smaller scales.  Many of the interesting models for modifications to gravity on large scales naturally produce different behavior on small scales, due to inherent non-linear \emph{screening mechanisms} which cause the gravitational force law to revert back to GR's predictions in high-density regions like the solar system.  This screening effect can extend to scales far beyond the solar system, depending on the model parameters. Indeed, for many models recently discussed in the literature \cite{Khoury:2003aq,Khoury:2003rn,Hu:2007nk,Nicolis:2008in,Hinterbichler:2010es,Brax:2010gi}, most of the virialized volume of massive dark matter halos can be screened, with a screening transition radius occurring near the outskirts of halos.

The transition regions (from screened to unscreened) are promising regions to search for deviations from GR. 
For example, several works have shown that the velocities of infalling objects around clusters are enhanced in modified gravity theories due to the stronger gravitational force, and the phase space around these clusters can provide important tests of gravity \cite{Zu:2013joa,Lam:2012by,Jain:2010ka}. In principle, these enhancements could be detected (for example) by comparing cluster lensing masses to dynamically-derived masses since the two masses depend differently on the metric potentials in many modified gravity models compared with GR. This would essentially correspond to a nonlinear version of the $E_G$ estimator used to test gravity on larger scales \cite{Zhang2007,Reyes2010}.
However, these measurements are not easy to make in clusters as the ratio between the dynamical and lensing mass is itself  scale dependent, and its determination requires precise measurements of redshifts through spectroscopic surveys and lensing mass reconstruction. 

The splashback feature in halo density profiles lies precisely in the outskirts of clusters, near the expected screening transition region.  Empirically, this feature is located where the slope of the density profile of matter around halos decreases sharply in a narrow, localized region, its value falling below the expected slope of $-3$ for NFW profiles before rising back to larger slopes in the two-halo regime \cite{Diemer:2014xya}. Theoretically, this feature is understood as the outer caustic of the dark matter halo density profile, where the most recently accreted material is at its first turnaround after infall \cite{AD14,Shi:2016lwp}.  The region outside splashback has material that has never been inside the halo, while the splashback radius itself forms the boundary of the ``virialized'' or multi-streaming region. This radius therefore is set by material that has not been in the halo for more than a dynamical time and therefore carries information about both the energy it falls in with as well as the rate at which the potential is growing.

In principle, the measurement of this radius only requires the density distribution around dark matter halos. As clusters are the most dominant structures in their environment, the feature is most pronounced in these objects. Attempts have been made recently to measure the feature around clusters of galaxies selected by the RedMaPPer algorithm \cite{Rykoff2014}. The splashback radius was measured in the projected distribution of galaxies around clusters in both SDSS data \cite{More16, Baxter:2017csy} and in DES Y1 \cite{Chang:2017hjt}. The feature was also measured in the projected density obtained from weak lensing in \cite{Chang:2017hjt}. Surprisingly, the measured radius of the minimum slope differs significantly from that expected from N-body simulations both in the case of lensing measurements and when measured using galaxy number densities. In particular the splashback feature is located at a radius about $20\%$ smaller than that expected from halos of the same mass in simulations, in both SDSS and DES-Y1. However, it was shown in \cite{Busch:2017wnu,Chang:2017hjt} that the cluster selection algorithm may itself shift the inferred location of splashback for optically selected samples. Therefore, whether or not the discrepancy from theory is a true physical effect still remains to be determined. 

In any event, the splashback feature is a signature in the spatial distribution of matter or galaxies that is related to the infall dynamics of recently accreted material. Its location is set simply by the collapse of matter shells under gravity in an expanding universe. In this paper we investigate whether the location of this feature is sensitive to the cosmological expansion, and the theory of gravity. We begin in section \ref{sec:1} by reviewing the analytical model set up in \cite{AD14} to predict the location of the feature and extending the analysis beyond dark energy as a cosmological constant. In the following sections we investigate how the location of the splashback radius will respond to modifications of gravity, we study its behavior in cluster mass halos in simulations of both nDGP and $f(R)$ theories as examples of Vainshtein and chameleon screening respectively.

\section{Splashback radius in CDM with Dark energy}
\label{sec:1}

As noted above, the splashback radius forms the boundary in configuration space between the multi-streaming region of the dark matter halo and the region of pure infall. The spherically averaged radius occurs near the first apocenter of particles or subhalos that have been accreted on to the host recently. This is the radius at which the accreted material is turning around in its orbit for the first time after infall. In spherical symmetry without substructure, the location of this feature can be simply predicted using the spherical collapse model, modified to include the effects of mass accretion on to the host halo \cite{AD14}. The location of the feature primarily depends on the accretion rate of the host halo and the redshift at which the halo forms. The model evaluates the overdensity within an infalling shell when it is turning around its first orbit. The predictions from the simple collapse model of \cite{AD14}  match the location of the feature in full $\Lambda$CDM N-body simulations \cite{AD14,Diemer:2014xya,Shi:2016lwp} quite well for a range of accretion rates and redshifts, and for profiles of stacked clusters in different mass ranges. This agreement implies that the location of the feature is robust to stacking over a wide range of cluster shapes, sizes and histories.  Note that this close agreement is found in the stacked profiles of ensembles of halos.  For individual halos, there can be significant deviations in the splashback surfaces from spherical symmetry due to substructure and triaxiality \cite{Mansfield2017,Diemer2017}.

\begin{figure}
    \includegraphics[width=0.55\textwidth, trim= 1.6cm 0cm 0in 0in,clip]{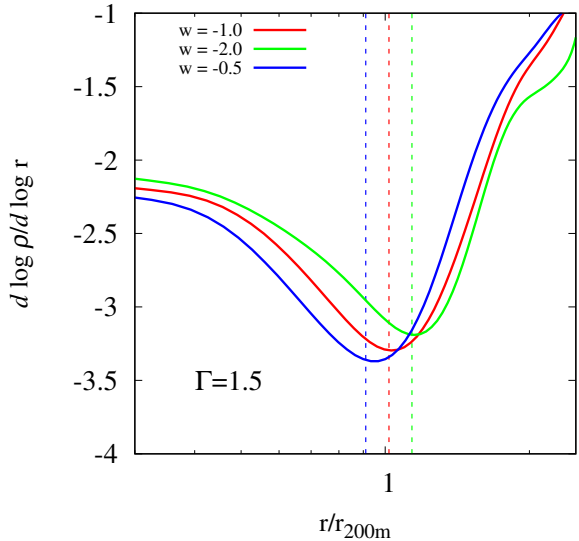}
	\caption{The logarithmic slope of the local density as a function of $r/r_{200}$ for stacked N-body halos at $z=0$ with virial mass $M_{\rm vir}=1- 4\times 10^{14}h^{-1} M_\odot$ and accretion rate $\Gamma=1.5$. Different colors correspond to different values of the equation of state parameter, $w$. Blue, red and green correspond to $w=-0.5,-1,$ and $-2$ respectively.  The vertical dashed lines show the expected position of splashback from the analytical model of \cite{AD14}. \label{fig:w_comp}
}
\end{figure}

The model for the location of the splashback feature in \cite{AD14} includes the effect of the cosmological constant. In general we expect dark energy to affect the location of the feature because it affects the background expansion rate of the universe.  
It is simple to modify the model to introduce a dark energy equation of state where, $P_{DE}=w\rho$  with $w\ne -1$. 
The background expansion rate is given by, 
\begin{equation}
\frac{\dot a}{a}=H_0\sqrt{\Omega_m a^{-3}+\Omega_{DE} a^{-3(1+w)}}
\end{equation}
and the equation of motion of a shell at radius $r$ becomes
\begin{equation}
\ddot r = -\frac{GM(r)}{r^2} - \frac{H_0^2}{2} \Omega_{DE} (1+3w) r^{-2-3w}.
\label{EOM2}
\end{equation}
These equations can be solved following \cite{AD14} to evaluate the location of splashback as a function of $w$, $\Omega_m$, and accretion rate $\Gamma= {\dd\ln M}/{\dd\ln a}$. To test the model we use N-body simulations of cosmologies with different equation of state parameters for dark energy. For this purpose we modified the background evolution equations in Gadget2 \cite{Gadget2} to allow for constant values of $w$ different from $-1$. We used $1024^3$ particles in a 1 Gpc $h^{-1}$ box. The softening length was chosen to be a quarter of a scale radius for a halo with 1000 particles, which is of the order of $L/(30N)$ \cite{Zhao:2008wd}. The cosmological parameters of the simulation were $\Omega_m=0.27$, $\Omega_{DE}=0.73$, $\Omega_b=0.0469$, and $h=0.7$. Fig.\ \ref{fig:w_comp} shows the comparison of our predictions from the toy model (eqn.\ \eqref{EOM2}) with the N-body simulations. We begin the simulations at an initial redshift of $z=49$, with initial conditions generated using N-GenIC\footnote{We have used Zeldovich approximation to generate initial conditions in this paper. As the initial redshift is fairly low a method like 2LPT is preferred (see \cite{Crocce:2006ve}). However, we find that the location of splashback for stacked halos in $\Lambda$CDM is in agreement with publicly available Multidark simulations (MDR1) \cite{Multidark} used in previous studies \cite{AD14}, therefore we do not change the initial condition generation method here.}  \cite{Gadget2}. Our model correctly captures the movement of splashback with changing $w$. We use $w=-0.5$ and $w=-2$, to amplify the effect of changing the EoS. 

The overdensity at splashback increases with increasing $w$. 
This behavior can be understood by considering again the rate at which the universe expands between turnaround and splashback. If $\Omega_{DE}=0.7$ today and $-1<w\le-1/3$, then dark energy domination begins at an earlier time than for a $w=-1$ universe. However at the time of turn around $\Omega_{DE}$ is higher than what it would have been for a $\Lambda$CDM universe, and therefore the background universe diluted faster than it would have for a $w=-1$ universe between turnaround and splashback, making the overdensity larger at splashback. The opposite is true for $w<-1$.

As is seen in Fig. \ref{fig:w_comp}, the splashback radius is sensitive to the equation of state parameter. However, to get differences in splashback larger than $10\%$, we need large deviations from $\Lambda$CDM that are already ruled out by observations. Percent level uncertainties on splashback measurement would be required to constrain dark energy at a level competitive with current bounds. Statistical errors on the splashback radius using the galaxy profile are already quite small, but systematic uncertainties are regarded as being significantly larger due to the cluster finder algorithm and other issues. Lensing measurements and the use of cluster finders that trace the mass distribution more closely are clear avenues for progress, but for these approaches statistical uncertainties will only reach the percent level in the next decade, with upcoming galaxy surveys (from LSST \citep{Abell:2009aa}, Euclid \citep{Laureijs:2011gra} and WFIRST \citep{Spergel:2013tha}) and CMB surveys (the Simons Observatory \citep{Simonssurvey} and CMB-S4 \citep{Abazajian:2016yjj}).

\section{Splashback in modified gravity models}

Modified gravity models have been invoked as an alternatives to dark energy to understand the large scale accelerated expansion of the universe. In these models, gravity is modified on large scales but on small scales, in higher density environments, general relativity must be restored to be consistent with the stringent observational tests of GR in the solar system. Most theories of interest therefore invoke \emph{screening mechanisms} to suppress these modifications in high density regions. Here we focus on the Hu-Sawicki $f(R)$ \cite{Hu:2007nk} and the Dvali-Gabadadze-Porrati (DGP) model \cite{Dvali:2000hr} that utilize two different classes of screening mechanisms, chameleon screening and Vainshtein screening, and see how they affect the splashback feature.

The modifications to GR can be parametrized by an enhancement of the gravitational constant in the unscreened region. The transition regions between the two regimes, screened and unscreened, often provide interesting scales for testing these theories. If the transition region lies in the outskirts of a halo, then accreted objects are in a region of enhanced gravity during the first infall but they may subsequently enter the screened region of the halo. We might expect that the varying gravitational field during the orbit of a particle may induce significant displacement of the splashback radius. 

In the following sections we briefly discuss the two main classes of model we choose to study the effect of modifications to GR on splashback.

\subsection{Chameleon screening \texorpdfstring{$f(R)$}{TEXT}}

One viable and well-studied model of modified gravity is the Hu-Sawicki $f(R)$ model \cite{Hu:2007nk}. $f(R)$ modifications replace the Ricci curvature $R$ in the Einstein-Hilbert action with a generic function thereof: 
\begin{equation}
S=\int d^4x \sqrt{-g}\frac{R+f(R)}{16 \pi G}.
\label{a}
\end{equation}
This gives dynamics to a third scalar polarization of the metric as well as the two tensor modes of GR. For this reason, $f(R)$ models can be recast as scalar-tensor theories with a fifth-force mediated by a scalar \cite{Brax:2008hh}. Screening in these models is achieved by a nonlinear coupling between the scalar field and matter, making the mass of the scalar field very high in dense regions thus reducing its Compton wavelength. This mechanism is known as chameleon screening. 
In terms of the $f(R)$ formalism, the additional degree of freedom is $f_R=df/dR$. The Compton wavelength of the field is given by $\lambda_{\rm C}^2={3 df_R/dR}$ and, in the absence of screening, the strength of gravity is enhanced by a factor of $4/3$ at distances within the Compton wavelength.

The \citet{husawicki} $f(R)$ model is given by,
\begin{equation}
f(R)=-m^2 \frac{c_1 (R/m^2)^n}{c_2(R/m^2)^n+1}
\label{hu_sawicki}
\end{equation} 
where $n$, $c_1$, and $c_2$ are model parameters, $c_1/c_2=6\Omega_\Lambda/\Omega_m$ to match the $\Lambda$CDM background evolution, and $m^2=\Omega_m H_0^2$. The model is parametrized by the background value of the derivative of the the field, $|f_{R0}|=-n c_1/c_2^2[3(1+4\Omega_\Lambda/\Omega_m)]^{-(n+1)}$ \cite{zhao11,Lombriser:2014dua}.

The coupled Poisson equations in the weak-field limit of Hu-Sawicki $f(R)$ gravity are:
\begin{align}
\nabla^2\Psi&=\frac{16\pi G a^2}{3}\delta\rho+\frac{a^2}{6}\delta R(f_R),
\label{fr1}\\
\nabla^2\delta f_R&=-\frac{a^2}{3}\left(\delta R(f_R)+8\pi G\delta \rho\right),
\label{fr2}
\end{align}
where $\Psi$ is the Newtonian potential, $\delta f_R=f_R(R)-f_R(\bar{R})$, $R$ is the Ricci scalar curvature, and $\delta R=R-\bar{R}$ and the barred quantities are the background values. When $\delta f_R=0$ there can be no source and from \eqref{fr2} we have $8\pi G\delta \rho=-\delta R(f_R)$. Plugging this into equation \eqref{fr1} one finds precisely the Poisson equation $\nabla^2\Psi = 4\pi G\delta \rho$. The other extreme limit is where $\delta R(f_R)\ll \delta\rho$ so that equation \eqref{fr1} gives $\nabla^2\Psi = 4\pi\times(4G/3)\delta \rho$ and the strength of gravity is enhanced by a factor of $4/3$. The former limit corresponds to the chameleon regime where gravity looks like GR and the latter regime is the unscreened regime. It can be shown that objects whose surface potentials $\Psi=GM/Rc^2<3f_{R0}/2$ 
will be unscreened whereas those with $\Psi>3f_{R0}/2$ will be in the chameleon regime. We refer the reader to \cite{Schmidt:2010jr,Lombriser:2014dua,Burrage:2017qrf} for more details on chameleon screening in $f(R)$ theories.

\subsection{nDGP and Vainshtein screening}

DGP is a braneworld model of the universe where we live on a four dimensional brane embedded in a five dimensional bulk. The decoupling (low energy) limit is described by general relativity and an extra scalar degree of freedom that mediates an additional gravitational-strength force. Physically, the scalar, or \emph{brane-bending mode}, describes the position of the brane in the 5D bulk. The screening is achieved by non-linear higher-derivative self-interactions of the scalar field\footnote{The structure of the higher-derivative terms are such that the equations of motion are second-order and therefore the Ostrogradski ghost is absent.}. The non-linearities are important near massive objects and result in a suppression of the sourced scalar field gradient compared with the Newtonian potential. Far away from the object the non-linearities are not important and the field gradient (additional scalar force) is comparable with the Newtonian one so that large deviations from GR are expected. The transition between the two regimes happens at the \emph{Vainshtein radius} and this screening mechanism is known as the Vainshtein mechanism. 
To study the effects on the outer density profile we choose the nDGP model, i.e. the normal branch of the DGP theory. Unlike sDGP, which is the self-accelerating branch of DGP, this has the same expansion history as $\Lambda$CDM, therefore we simulate this model to disentangle the effects of modified gravity.

The Friedmann equation in the nDGP model is given by,
\begin{equation}
H^2=-\frac{H_0}{r_c}+\frac{8\pi G \rho}{3}
\label{dgp_eqn}
\end{equation}
where $r_c$ (the \emph{crossover scale}) is the scale at which gravity transitions from being 5-dimensional on large scales to 4-dimensional on small scales. The background evolution is tuned to match the $\Lambda$CDM expansion. The Poisson equation and the scalar field equation are given by,

\begin{align}
\nabla^2\Psi_N+\frac{1}{2}\nabla^2\phi&=\nabla^2\Psi\label{dgp1}\\
\nabla^2\phi+\frac{r_c^2}{3\beta(a)a^2}\left[(\nabla^2\phi)^2-(\nabla_i\nabla_j\phi)(\nabla^i\nabla^j\phi)\right]&=\frac{8\pi G a^2}{3\beta(a)}\rho\delta,
\label{eq:dgp2}
\end{align}
where $\phi$ is the scalar field $\Psi$ is the Newtonian potential, $\nabla^2\Psi_N=4\pi G \bar{\rho}\delta$ is the Newtonian potential in GR, $\delta$ is the density contrast, and the parameter $\beta(a)$ is given by,
\begin{equation}
\beta(a)=1+2 H r_c\left(1+\frac{\dot{H}}{3 H^2}\right).
\label{eq:dgp_beta}
\end{equation} 
As $r_c$ becomes large the modifications to gravity become weaker and the Vainshtein screening mechanism becomes more efficient. In the spherically symmetric case the Vainshtein radius is given by
\begin{equation}
r_\star= \left(\frac{16 G M(r) r_c^2}{9 \beta^2}\right)^{1/3}.
\label{eq:v_rad}
\end{equation}
Here, $M(r)$ is the mass enclosed inside radius $r$, which can be found by integrating the NFW profile for a single halo.

\subsection{Evolution of shells in nDGP: worked example}

The Poisson equation in both models described above becomes highly nonlinear, making non-linear evolution difficult to solve analytically, however one can gain sufficient insight into the dynamics of shells in these models of enhanced gravity by making simplifying assumptions.  

In this section we work out exactly the dynamics of particle shells in the nDGP model. This example is demonstrative of the differences that making gravity stronger can bring to the evolution of shells and therefore halos. We highlight the different aspects of evolution that will be altered in this scenario and in the following section we perform an extensive analysis of simulated cosmologies with insights from the worked example.

The aim is to calculate the orbit of an infalling shell in nDGP. The force term in the equation of motion of the shell in $\Lambda$CDM is modified by the force mediated by the scalar field $F_\phi$.

\begin{equation}
\ddot r = -\frac{GM}{r^2} + \frac{\Lambda\,c^2}{3} r + F_\phi,
\label{EOM}\end{equation}
where $\phi$ is the new gravitational scalar that satisfies equation \eqref{eq:dgp2}.

\subsubsection{Modification to the force law}

Taking the perturbed metric (in the Newtonian gauge) to be 
\begin{equation}
\mathrm{d}s^2=-(1+2\Psi(t,x))\mathrm{d}t^2+a^2(t)(1+2\Phi(t,x))\mathrm{d}x^2,
\end{equation}
the equation of motion for the potentials $\Phi$ and $\Psi$ is not $\Phi=-\Psi$ as it is in GR but is modified by the scalar so that one has
\begin{equation}
\Psi= \Psi_N+\frac{1}{2} \phi,
\end{equation}
\begin{equation}
\Phi= -\Psi_N+\frac{1}{2} \phi,
\end{equation}
where, $\Psi_N$ is the Newtonian potential and $\phi$ is a scalar field. Together they contribute to the Poisson equation,
\begin{equation}
\nabla^2\Psi=\nabla^2\Psi_N+\frac{1}{2}\nabla^2\phi
\label{poisson}
\end{equation}
For a spherically symmetric mass distribution equation \eqref{eq:dgp2} becomes,
\begin{equation}
\frac{1}{r^2}\frac{d}{dr}\left[r^2\phi'+\frac{2r_c^3}{3}(r\phi'^2)\right]=\frac{8\pi G}{3\beta(a)}\rho
\end{equation}
where the prime denotes derivative with respect to $r$. This equation can be multiplied by $r^2$ and integrated to give,
\begin{equation}
\frac{2r_c^2\phi'^2}{3r}+\phi'-\frac{2}{3\beta(a)}\frac{GM(r)}{r^2}=0
\label{quad}
\end{equation}
where the mass, $M$ is the enclosed mass at radius $r$ given by,
\begin{equation}
M(r)=4\pi\int_{0}^{r}\rho_m(r)r^2dr.
\end{equation}
Equation \eqref{quad} is a quadratic in $\phi'$ and solving we find
\begin{equation}
\phi'=-F_\phi=\frac{4}{3\beta(a)}\frac{GM(r)}{r^2}g(r/r_\ast)
\end{equation}
where, $g(\zeta)=\zeta^3(\sqrt{1+\zeta^{-3}}-1)$, and
$r_\ast$ is the (radially-dependent) Vainshtein radius of the system as described in equation \eqref{eq:v_rad}. 

The function $\Delta G_{DGP}=2G/(3\beta) ~ g(r/r_\ast)$ acts like a correction to the gravitational constant. When $r>>r_\ast$, $\Delta G_{DGP}=1/(3\beta)$ and, including the Newtonian force, the effective Newtonian constant $G_{\rm eff}=4 G/(3\beta)$.

$F_\phi$ gives the modification to the force from $\Lambda$CDM. Following \cite{AD14} we solve equation \ref{eq:dgp2} with an NFW profile for the mass, which is growing with time as
\begin{equation}
M(r)=M_{tot}\frac{f_{\rm NFW}(r/r_s)}{f_{\rm NFW}(c)}; ~~ M_{\rm tot} \propto a^\Gamma
\label{nfw}
\end{equation} 
where $\Gamma={\dd\ln M}/{\dd\ln a}$ is the accretion rate of the halo.

\subsubsection{Background evolution}

To find the overdensity at splashback we need to define the background evolution. For the normal branch of DGP the Friedmann equation is given by, 

\begin{figure}
	\includegraphics[width=0.45\textwidth, trim= 0.5in 0cm 0in 0in,clip]
    {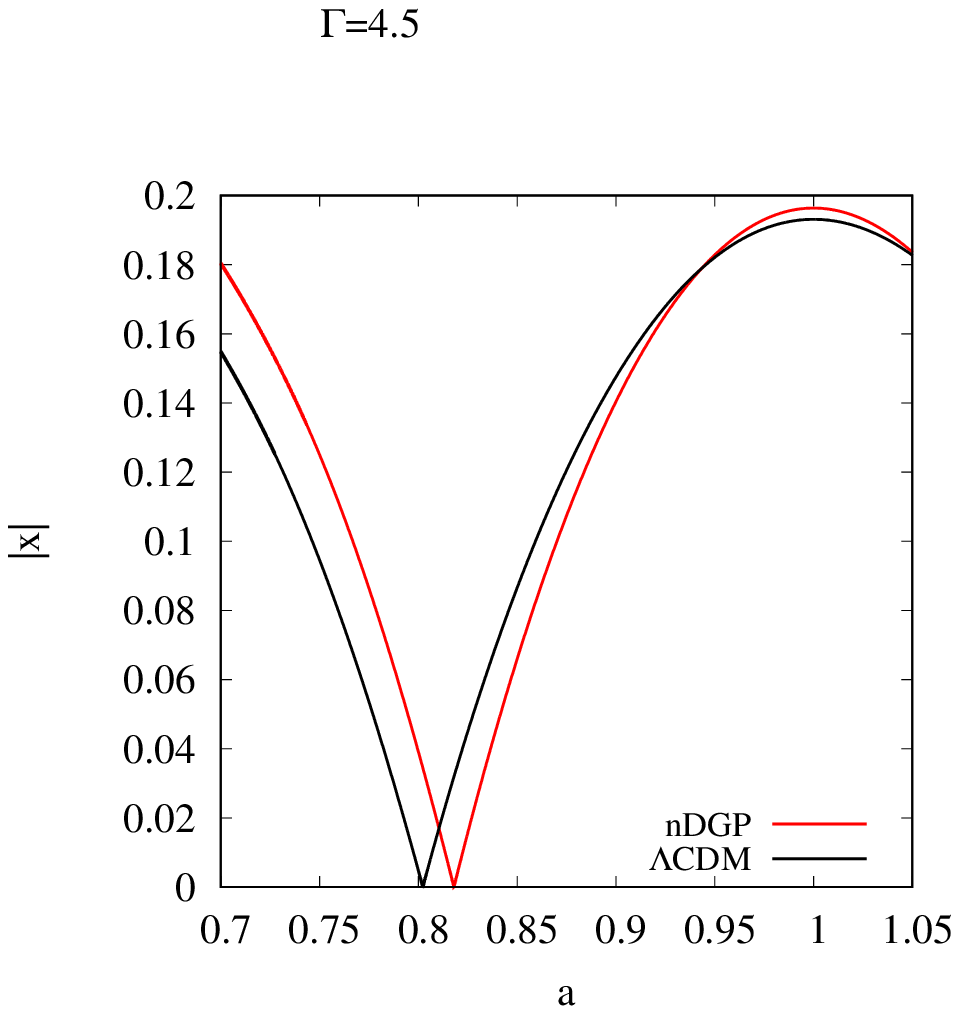}
	\includegraphics[width=0.45\textwidth, trim= 0.5in 0cm 0in 0in,clip]
    {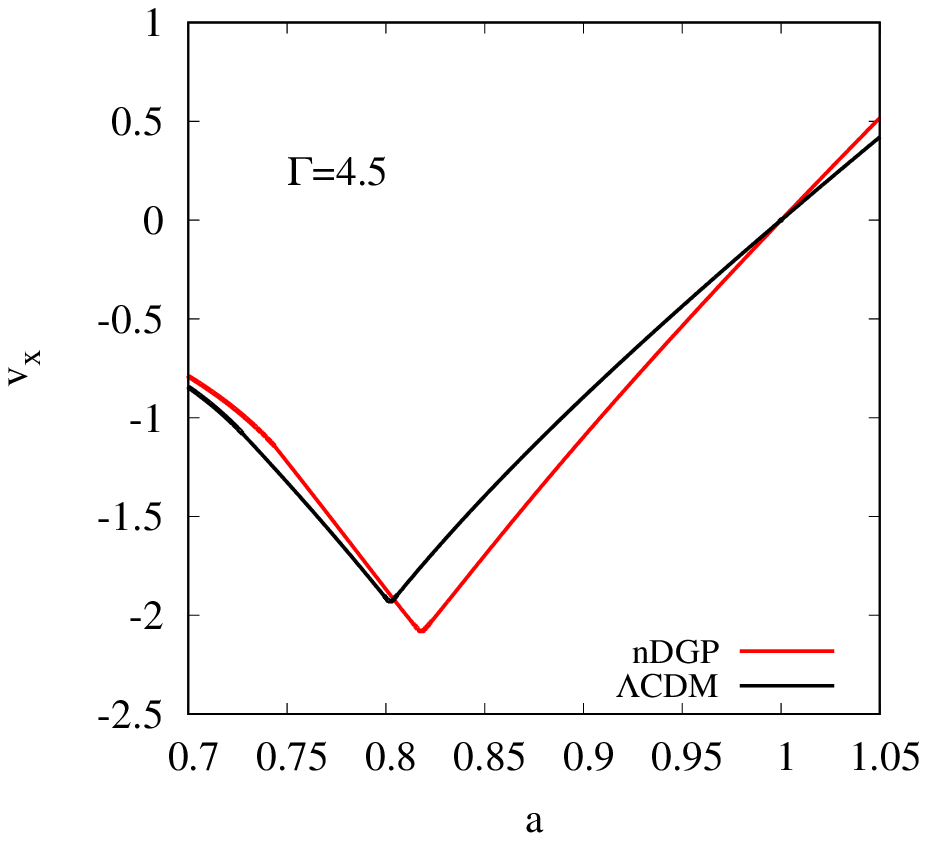}	 
	\caption{
    {\it Top:} The evolution of the radius of collapsing shells in our model for nDGP with $H_0 r_c=0.2$ and $\Lambda$CDM, with the initial conditions  adjusted such that the shells splash back at $a=1$ in both cases. {\it Bottom:} The velocity of the shell with the Cartesian $x-$axis as the radial direction.\label{splash_modg}}
\end{figure}

\begin{eqnarray}
\frac{H(a)}{H_0} =\sqrt{\Omega_m a^{-3}+\Omega_{DE}(a)+\Omega_{rc}}-\sqrt{\Omega_{rc}},
\label{background}
\end{eqnarray}
where $\Omega_{rc} \equiv 1/(4H_0^2r_c^2)$, $\Omega_m=\rho_m/\rho_{cr,0}$ is the average matter density ratio, and $\Omega_{DE}(a)=\rho_{DE}(a)/\rho_{cr,0}$ is the ratio of the smooth dark energy component to the critical density. The evolution of the background can be matched exactly to $\Lambda$CDM by equation \eqref{background} to the Friedmann equation in GR with $\Omega_\Lambda=1-\Omega_m$. This basically means we have a dark energy component with an effective, time-varying density given by, 

\begin{equation}
	\Omega_{DE}(a)=\left[\Omega_\Lambda+2\Omega_{r_c}\left(\sqrt{\Omega_m/\Omega_{rc}a^{-3}+1}-1\right)\right].
	\label{denergy}
\end{equation}

\subsubsection{Initialization of the shell: spherical collapse in nDGP}

In the model for splashback we follow the time evolution of a shell in an accreting dark matter halo in an expanding universe. We use spherical collapse with a constant interior mass until the shell reaches half the turnaround radius, thereafter the mass profile is given by the time evolving NFW profile specified in equation \eqref{nfw}. 
We can treat modified gravity models in a similar manner. We can solve the spherical collapse model with constant interior mass for nDGP to get the initial evolution of the shell before entering the ``virialized" region.

Following \cite{Schmidt09} we can work out the spherical collapse solution for nDGP. 
We consider a constant overdensity $\delta_i$ in a region of radius physical $R_i$ in the early universe. The mass enclosed within the shell is conserved in time and is is given by,
\begin{equation}
M_i=M=\frac{4\pi}{3}R^3\bar{\rho}_{m}(1+\delta),
\label{scoll}
\end{equation}
where $R$ is the radius of the shell, $\delta$ is the enclosed overdensity, and $\rho_m$ is the background matter density at current time. 

The equation for the evolution of a density perturbation in the early universe is given by,
\begin{equation}
\ddot{\delta}-\frac{4}{3}\frac{\dot{\delta}^2}{1+\delta}+2H\dot{\delta}=(1+\delta)\nabla^2\Psi.
\label{pert}
\end{equation}
\begin{figure}
	\centering
	\includegraphics[width=0.45\textwidth, trim= 1.5cm 0.1cm 0in 0in,clip]{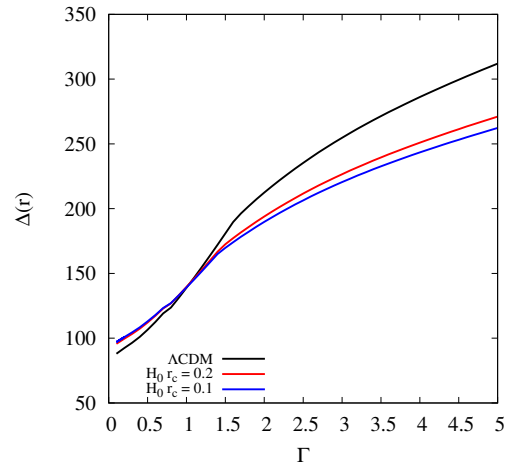}
	\caption{The overdensity at the splashback radius as a function of accretion rate for the three models shown in Figure 2. A higher overdensity corresponds to a smaller splashback radius. Differences between the models are evident at higher accretion rates. \label{fig:delta}}
\end{figure}%
$\nabla^2\Psi$ is given by the Poisson equation \eqref{poisson}. Using equation \eqref{scoll} we may replace $\delta$ in \eqref{pert} and rewrite it as,
\begin{align}
\frac{\ddot{R}}{R} &= H^2+\dot{H}-\frac{1}{3}\nabla^2\Psi \nonumber \\
                   &= -\frac{4\pi G}{3}(\bar{\rho}+\rho_{DE})-\frac{4\pi G}{3}\left(1+\frac{2}{3\beta}g(R/R_\ast)\right)\delta\rho
\end{align}
We can then solve the equation numerically. Using the convention in \cite{Schmidt09, Schmidt08} we define,
\begin{equation}
x=\frac{R}{R_i}-\frac{a}{a_i}.
\end{equation}
The evolution equation then becomes, with the time coordinate $\ln a$,
\begin{eqnarray}
\ddot{x}=&&-\frac{\dot{H}}{H}\dot{x}+\left(1+\frac{\dot{H}}{H}\right)x\\
&&-\frac{\Omega_m H_0^2a^{-3}}{2H^2}\left[1+\frac{2}{3\beta}g(R/R_\ast)\right](x+a/a_i)\delta,
\end{eqnarray}
where $\delta(x,a)=(1+\delta_i)\left(a_i x/a+1\right)^{-3}-1$. We begin with $x=0$ and $\dot{x}=-\delta_i/3$. We adjust $\delta_i$ such that splashback occurs at $a=1$. 

Figure \ref{splash_modg} shows the evolution of a shell that is at the first apocenter of its orbit, or splashback,  today. One can see that the splashback radius can move to larger radii in the case of modified gravity with $H_0 r_c =0.2$ for high accretion rates. Fig.\ \ref{fig:delta} shows the enclosed overdensity as a function of accretion rate for $H_0 r_c= 0.2$ and $H_0 r_c=0.1$.  Since high mass, cluster sized halos are typically accreting rapidly, we can hope to detect the percent level differences in splashback radius by observing massive objects. 
The bottom panel of figure \ref{splash_modg} shows the velocity of the infalling shells, which is significantly different in the two cases. The stronger gravity leads to shells in nDGP infall with a higher velocity than its counterpart in $\Lambda$CDM. As dynamical friction goes as $1/v^2$, differences in the trajectory of massive satellites in the two cosmologies are likely to introduce changes in the splashback radius.

\section{Splashback radius in simulations of modified gravity}

\subsection{Splashback for particles}

With insights from the toy model for the spherical case, we study the full non-linear growth of structure using simulations of the different modified gravity models. As mentioned, the Poisson equation in modified gravity models become highly nonlinear, making these models difficult and time-consuming to solve numerically. ECOSMOG-V and ECOSMOG-fR \cite{Bli12V,Bli13V} are state of the art codes to simulate cosmologies with modified gravity. They are adaptive mesh refinement (AMR) codes based on RAMSES, that solve for the scalar field $\phi$ using the Newton-Gauss Siedel multigrid relaxation method. For a detailed review of the code performance refer to \citet{Bli13V,Bli12V}. 

We run and analyze a suite of N-body simulations both in $f(R)$ gravity and in nDGP. Firstly, to maximize the effects of modifications we simulate the nDGP models of the worked example with $512^3$ particles on a $400$ Mpc $h^{-1}$ length box. We begin at an initial redshift $z=49$. We run these with the best-fit Planck cosmology parameters, $\Omega_m=0.301$, $\Omega_\Lambda=0.699$, $h=0.677$, and $n_s=0.9611$. We simulate the nDGP model with $H_0 r_c = 0.2$, where the crossover scales is 600 Mpc $h^{-1}$. As the growth of structure in DGP is different from $\Lambda$CDM, we normalize the initial conditions of our simulations, such that $\sigma_8$ is the same for both the nDGP and the corresponding $\Lambda$CDM case. 

\begin{figure*}
	\centering    \includegraphics[width=1.1\textwidth, trim= 0cm 4.55cm 0cm 0.2cm,clip]{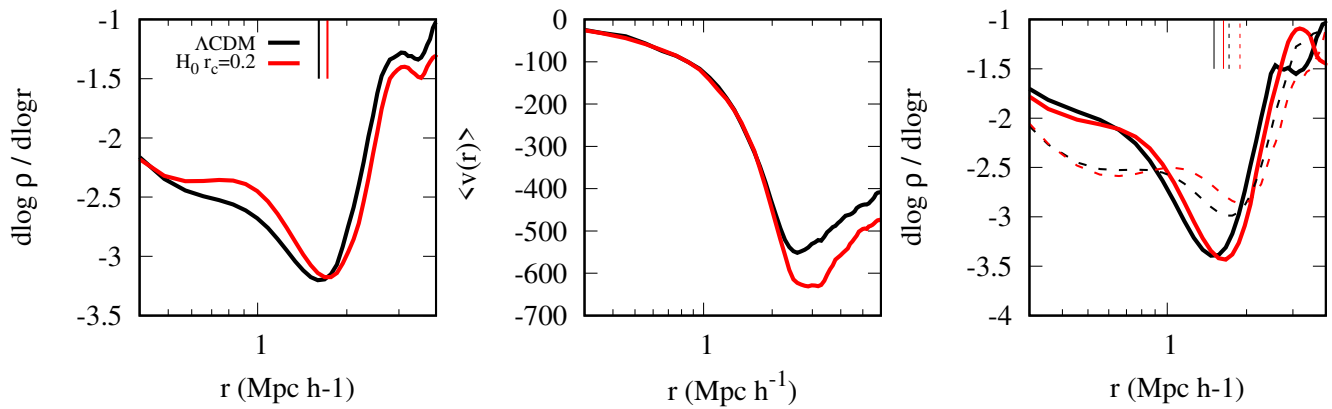}
	\caption{{\it Left:} Slope of the stacked density profile of  halos in the mass bin from $M_{vir}=1$--$4 \times 10^{14}M_\odot h^{-1}$ in nDGP and $\Lambda$CDM. {\it Middle:}  Stacked average radial velocity in km/s as a function of radius. {\it Right:} Comparison between the slope of the stacked density profile of halos in nDGP and $\Lambda$CDM in different concentration bins. The dashed curve corresponds to a mean concentration $c=6$ while the solid curves are for lower concentration halos with mean $c=3$. The vertical lines correspond to the position of the minimum slope. The difference between the models is about 10 percent or larger. }
    \label{fig:simnDGP}
\end{figure*}

Fig.\ \ref{fig:simnDGP} shows the comparison between the slopes of the density profile  for cluster mass halos in $\Lambda$CDM and nDGP with $H_0 r_c=0.2$. The value of $r_c=$ 600 Mpc $h^{-1}$, which corresponds to $H_0 r_c=0.2$, was chosen so that the Vainshtein radius for cluster sized halos lies in the splashback region. Taking the crossover scale $r_c\sim\mathcal{O}(6000\textrm{ Mpc})$ so that $H_0r_c\sim\mathcal{O}(1)$ results in cosmic acceleration driven by the modifications of gravity i.e. DGP (or galileons) play the role of dark energy. This would correspond to the so-called \textit{sDGP} scenario. This scenario is strongly ruled out since it has ghosts in the spectrum (so that space-time is unstable) \cite{Luty:2003vm,Nicolis:2004qq,Koyama:2005tx,Charmousis:2006pn,Koyama:2007zz}. In the case of galileon dark energy, self-accelerating models are either ruled out by the bounds on the difference between the speed of light and gravitational waves coming from GW170178/GRB 170817A \cite{Sakstein:2017xjx,Baker:2017hug,Creminelli:2017sry,Ezquiaga:2017ekz} or are in strong tension with cosmological observations \cite{Renk:2017rzu, Barreira:2014jha}.  The values of $H_0 r_c$ considered above correspond to the strongest small-scale modifications of gravity that are not yet ruled out by other probes \citep{Sakstein:2017bws,Sakstein:2017pqi}.
The red and blue curves show the profiles for the nDGP models. The top panel shows the splashback radius for stacked cluster mass halos with masses $1\le M \le 4\times 10^{14} M_\odot h^{-1}$. As expected from the model the location of the minimum of the density moves to a higher value compared with $\Lambda$CDM. The bottom panel shows the average radial velocity as a function of radius for the same sample halos. Particles that are at splashback today fall in with velocities that can be considerably higher than their counterparts in GR as can be seen from the comparison of the average radial velocities in the outskirts of the cluster (where gravity is unscreened). The larger velocities cause the particles in nDGP to splash back to larger radii compared with GR due to their higher energies. 

As the splashback radius depends on the growth history of halos, stacking halos with similar accretion rates or concentrations often makes the feature more prominent. The right panel of Fig.\ \ref{fig:simnDGP} shows the slope of the density profiles of halos in the same mass bin but split based on their concentrations, $c=R_{\rm vir}/r_s$, where $R_{\rm vir}$ is the virial radius and $r_s$ is the scale radius in the NFW fit to the profiles. The difference in the location of the feature then becomes more significant, of the order of 10$\%$.

Apart from the cases mentioned above we also consider models that are currently interesting in the cosmological context. For $f(R)$ gravity we look at models with $f_{R0}= 10^{-5}$ and $10^{-6}$, referred to as F5 and F6 henceforth. We also consider nDGP cosmologies with $H_0 r_c=1$, referred to as N1. The fiducial boxes are of length $1024$ Mpc$h^{-1}$ containing $1024^3$ particles. In these simulations the expansion is matched to a universe with $\Omega_m=0.281$, $\Omega_\Lambda=0.719$ and $h_0=0.697$.  We find that these modifications are weak enough that they do not produce an observable effect in the position of the particle splashback. When split based on concentration within the given mass bin, however,  differences of the order of a few percent in the slope profile begin to appear.

\subsection{Splashback for Subhalos}

The distribution of dark matter can be traced in observations using lensing. However the distribution of substructure in halos is also expected to trace the overall matter distribution. Galaxies reside in dark matter halos that are captured by the host and orbit within it as substructure. Therefore apart from tracing the matter distribution with lensing one may also look at the distribution of galaxies or substructure to measure the feature in its density profile, as in Refs.\ \citep{More16,Chang:2017hjt}.

Unlike dark matter particles, however, massive subhalos experience dynamical friction when moving through an ambient density of background particles \citep{Chandrasekhar1949,BinneyTremaine}. The force from dynamical friction acts like a drag on subhalos, decreasing their orbital apocenters relative to those of dark matter particles \cite{AD16}. The deceleration due to dynamical friction is given by
\begin{equation}
\frac{d\bm{v}}{dt} \propto -\frac{G^2 M \rho}{v^3} \bm{v} f(v/\sqrt{2}\sigma),
\label{DF}
\end{equation}
where $M$ is the mass of the object moving with a velocity $v$ through an ambient medium of lower mass particles, $\rho(r)$ is the density of the medium, and $\sigma$ is the ambient velocity dispersion. The function $f(X)=\textrm{erf}(X^2)-\frac{2X}{\sqrt{\pi}}e^{-X^2}$.

The timescales for dynamical friction depend on the ratio of the masses of the subhalo and the host, and become relevant, i.e. of the order of Hubble time or shorter when $M_{\rm sub}/M_{\rm host}>0.01$. Therefore, the lowest mass subhalos in massive clusters have a splashback radius similar to that of particles, while higher mass subhalos splashback at radii smaller than the particles. The splashback radius of subhalos or galaxies in observations can therefore act as a direct probe of dynamical friction.

\begin{figure*}
	\includegraphics[width=0.44\textwidth, trim= 1cm 0cm 0cm 0.4cm,clip]{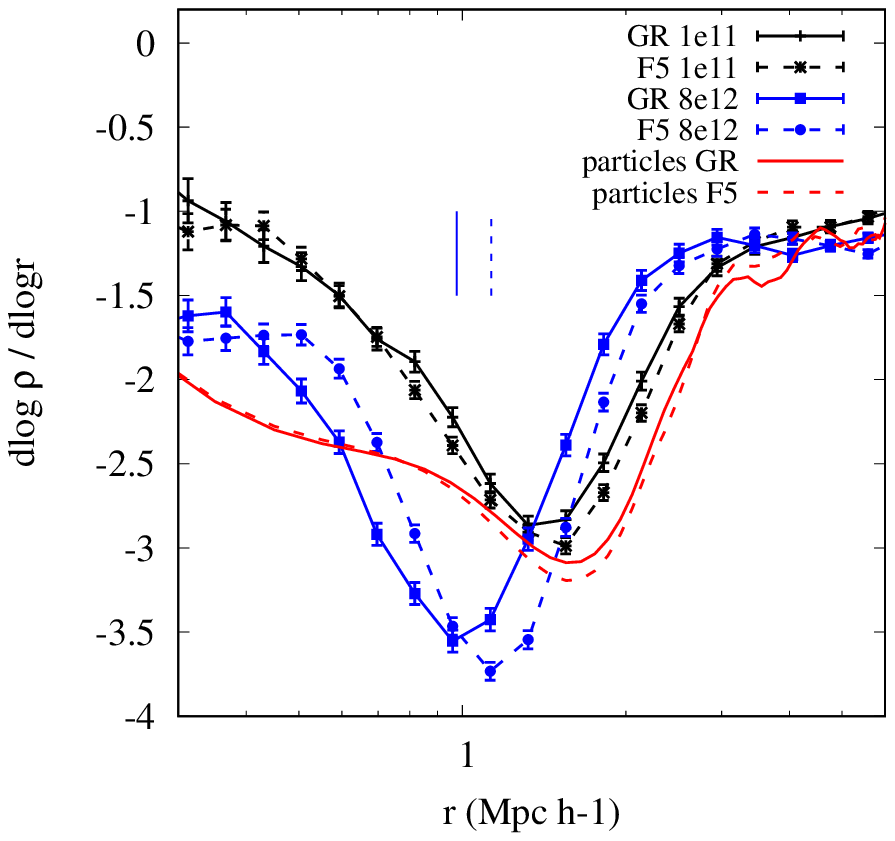}     \includegraphics[width=0.45\textwidth, trim= 1cm 0.3cm 0cm .2cm,clip]{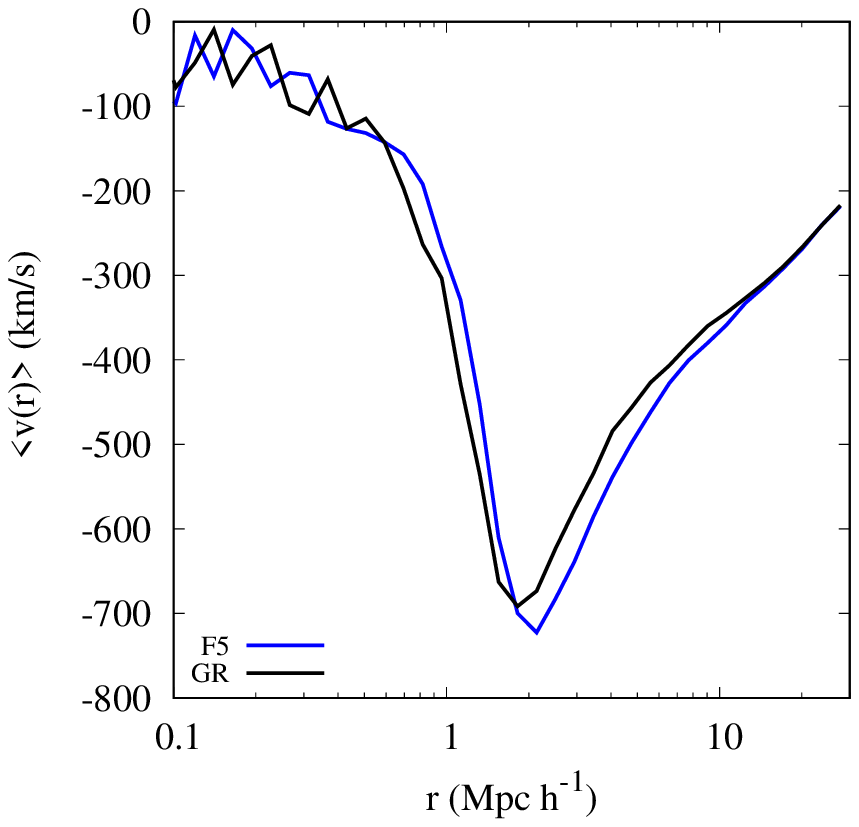}
    	\caption{{\it Left:} Comparison of splashback radius in $f(R)$ gravity with $f_R=10^{-5}$ (F5; dashed curves) and GR (solid). As indicated in the legend, results for the particles and subhalos with two mean masses are shown. The vertical lines correspond to the splashback radius for subhalos with $M_{\rm peak}>8\times10^{12} M_\odot h^{-1}$ in F5 (dashed) and GR (solid). These subhalos are affected by dynamical friction and respond differently in the two models. {\it Right:} Average radial velocity for subhalos of mass, $M_{\rm peak}>8\times10^{12} M_\odot h^{-1}$(right) in F5 (blue) and GR (black) around clusters of mass $M_{\rm vir}=1$--$4\times10^{14} M_\odot h^{-1}$. \label{fig:frgr}}
\end{figure*}

\begin{figure}	 
    \includegraphics[width=0.45\textwidth, trim= 1cm 0cm 0cm 0cm,clip]{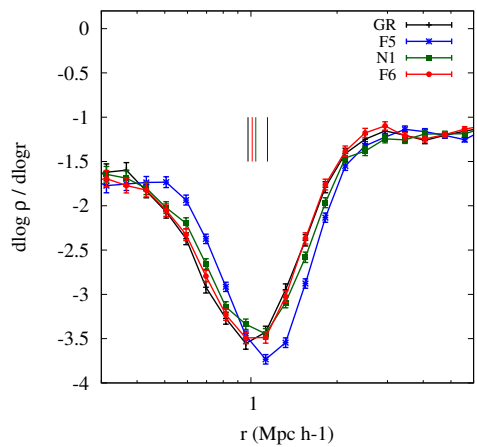}
	\caption{The effect of gravity theories on dynamical friction and splashback. The profile of subhalos with mass $M_{\rm peak} > 8\times 10^{12} M_\odot h^{-1}$ around clusters is shown for four different modified gravity models (vertical lines correspond to the splashback radius as in earlier figures). The signature is largest for the $f(R)$ model shown in Figure \ref{fig:frgr}. \label{fig:allsub}}
\end{figure}

We examine the distribution of subhalos around hosts of mass $M_{\rm vir} = 1-4\times 10^{14} M_\odot h^{-1}$ in the different modified gravity simulations. To learn about properties of very low mass subhaloes we ideally require high resolution simulations. However we only require the positions and velocities of the subhalos for our analysis of the splashback radius and no internal properties like accurate profiles of the subhalos themselves. For F5 we have two sets of simulations, one with 1024 Mpc $h^{-1}$ and $1024^3$ particles and another with higher resolution of 450 Mpc $h^{-1}$ with $1024^3$ particles. Fig \ref{fig:frgr} shows the splashback radius for subhalos in F5 and GR for subhalos of mass $1\times 10^{11} M_\odot h^{-1}$ and $8\times 10^{12}~ M_\odot h^{-1}$. We select subhalos based on their peak mass, $M_{\rm peak}$, measured from their merger histories, rather than their present mass.  The splashback radius for lower mass subhalos was computed using the higher resolution simulation. For comparison, the particle splashback for both the models is also shown in red. The low mass subhalos with $M_{\rm peak}>1\times 10^{11} M_\odot h^{-1}$ have a splashback radius that is close to the splashback radius for particles in both GR and F5, whereas the higher mass subhalos with $M_{\rm peak} > 8 \times 10^{12}$ have a smaller splashback radius than particles due to dynamical friction. However, the higher mass subhalos in $f(R)$ gravity have a  larger splashback radius in comparison to their counterparts in GR. In fact it appears that the effect of dynamical friction is suppressed in F5. The difference between the particle splashback and subhalo splashback for subhalos with mass $8\times 10^{12} M_\odot h^{-1}$ is smaller than it is for GR.  Fig.\ \ref{fig:allsub} shows the splashback for $M_{\rm peak} > 8\times 10^{12} M_\odot h^{-1}$ subhalos in all the different models. N1 and F5 both show evidence for reduced dynamical friction. 

This behavior can be explained by examining the infall velocities of subhalos into hosts. The force of dynamical friction depends on the relative velocity between a satellite and its hots; the higher the velocity, the lower the drag from friction.  Due to the enhanced gravity in the outskirts of the cluster mass halos, the infall velocities of particles and subhalos are enhanced in the stronger gravity models, which should lead to a suppression of dynamical friction. The right panel of Fig.\ \ref{fig:frgr} shows the mean radial velocity of the high mass subhalos in GR and F5, and, as can be seen, subhalos in the outskirts of clusters have higher infall velocities in F5 than in GR on average. Consequently on entry into the host, subhalos in F5 feel less frictional force than their counterparts in GR, and therefore splash back to a larger distance at apocenter. It should be noted that while low mass subhalos and particles also fall in with higher velocities, the dynamical friction timescales are much longer for low $M_{\rm sub} / M_{\rm host}$ and do not affect their dynamics significantly.

\section{Discussion}

In this paper we have studied  the sensitivity of the splashback radius to gravity and the nature of dark energy. We used an analytical model for the spherical evolution of matter falling onto cluster halos, which has previously been shown to reproduce splashback measurements in N-body simulations \cite{AD14}. We extended the model to account for evolving dark energy and modified gravity models with Vainshtein screening. We also studied a second gravity model, Hu-Sawicki $f(R)$, which screens using the chameleon mechanism.  

Firstly, we demonstrated that the location of the splashback feature is sensitive to the equation of state parameter $w$ of dark energy. This effect is primarily due to the change in the universe's background expansion rate. Precise measurements of splashback radius will be required to constrain cosmologically viable values of $w$, such that changes of the order of a few percent may be detected.

We found that modifications to general relativity could significantly alter the dynamics of infalling matter and lead to observable changes in the location of splashback in both the mass and subhalo profiles (Figs. 4 and 5). We studied splashback in two gravity models that exhibit screening mechanisms: nDGP, which exhibits the Vainshtein mechanism, and $f(R)$ gravity, which utilizes the chameleon mechanism. For nDGP we found that for small values of the crossover scale (of order 500 Mpc $h^{-1}$), the splashback radius is significantly larger than in $\Lambda$CDM. Stacking on concentration to group together halos of similar assembly histories makes the differences in splashback radius in the clusters more pronounced; it can be larger by about $10\%$ compared with GR. Weak lensing profiles around cluster mass halos would test modified gravity in this regime\footnote{For part of the parameter space in DGP/galileon ($r_c\lsim\mathcal{O}(\textrm{Gpc})$) that is interesting for cosmic acceleration we found that the splashback radius is nearly identical to GR, i.e. the particle trajectories within the halos are not modified on average. In any case, such models are in strong conflict with other observations. Our study may thus be regarded as illustrative of the effects of modified gravity; the quantitative signature of models developed in the future will need to be calculated for each model.}. 
%

Along with particles we also studied the splashback feature of subhalos in N-body simulations (Fig. 5). In GR, massive subhalos are expected to have a smaller splashback radius than particles owing to loss of orbital energy due to dynamical friction. We find that in $f(R)$ gravity when $f_{R0}>10^{-5}$, the difference between the splashback radius for particles and subhalos with mass $M_{sub}>8\times10^{12} M_\odot h^{-1}$ is reduced. This is due to the enhanced infall velocities from the gravitational force, which makes dynamical friction  less effective in clusters compared with GR. Comparisons between particle splashback measured from weak lensing and galaxy splashback in clusters can therefore be used as a probe for modified gravity models. Since galaxy luminosities are  correlated with their subhalo masses, the variation of the  splashback radius with galaxy magnitude should generically be weakened in such gravity theories. 

Beyond galaxy clusters, the effect of dynamical friction is important to understand the distribution of satellite galaxies in host halos of all masses. In addition, for $f(R)$ theories with $f_{R0} < 10^{-5}$, while cluster mass halos are completely screened, lower mass halos may remain partially unscreened in the outskirts, leading to a similar enhancement of velocity that should imprint its signature on the splashback radius. Due to resolution limits of our simulations we do not look into this regime; future studies with higher resolution simulations of modified gravity will be required to understand the effects of reduced dynamical friction on satellite distributions.

Observationally, splashback in galaxy clusters has been detected in both the galaxy and lensing profiles \citep{More16,Baxter:2017csy,Chang:2017hjt}. 
In current measurements of splashback in the galaxy profile, there appear to be systematic uncertainties at the 10 percent level, possibly due to the optical cluster detection approach. With cluster samples selected using proxies that are closer to a mass selection \cite{tae2018inprep} these issues may be resolved. If some of the current results on the splashback radius being smaller than expected hold, they could constrain gravity theories tightly as they generically predict larger splashback radii. 
A second feature of current measurements is that the splashback radius  shows no significant trend with galaxy magnitude for high richness bins, i.e. for cluster with mass $M>10^{14} M_\odot h^{-1}$.  For lower mass group-sized objects, there is evidence for movement of the location of splashback to smaller radii for bright galaxies \cite{AD16}. 
Pursuing the signatures of dynamical friction for galaxies of different observed magnitudes could have interesting implications for modified gravity. 

Future surveys like LSST \citep{Abell:2009aa}, and complete data from the ongoing DES \citep{DESsurvey}, KiDS \citep{Kidssurvey} and Subaru/HSC surveys \citep{Subarusurvey}, will achieve powerful statistical accuracy on measurements of splashback. In the results we have presented, variations due to modified gravity range from a few to ten percent. The splashback radius measured from the number density profile of galaxies will be constrained at the percent level in the coming years. Even from lensing, the current uncertainty of under 15 percent from the DES Year 1 measurement \cite{Chang:2017hjt}, could improve to 5 percent with the full DES survey, through a combination of increased sky coverage and improved methodology. With LSST, an additional factor of 3-4 in survey area and number density of source galaxies will lead to statistical uncertainty of 1-2 percent. Thus, if systematics can be controlled at the same level (clearly a major caveat) measurements with future surveys can distinguish even the more subtle model variations presented here. Moreover, the full shape of the density profile contains information beyond the location of splashback; we have not attempted to quantify this signal. Future spectroscopic surveys like DESI \citep{Aghamousa:2016zmz} will help constrain the relationship between stellar mass and halo mass better, helping with the interpretation of any trend of splashback with galaxy magnitudes.

\medskip

\begin{acknowledgments}
We thank Andrey Kravtsov, Surhud More, Eric Baxter, Chihway Chang, Tae-Hyeon Shin, Arka Banerjee, Benedickt Diemer and Fabian Schmidt for helpful discussions. 
This work was supported by NASA under grant HST-AR-14291.001-A from the Space Telescope Science Institute, which is operated by the Association of Universities for Research in Astronomy, Inc., under NASA contract NAS 5-26555. JS is supported by funds provided to the Center for Particle Cosmology by the University of Pennsylvania. BJ is supported in part by the US Department of Energy grant desc0007901. Research at Perimeter Institute is supported by the Government of Canada through Industry Canada and by the Province of Ontario through the Ministry of Research \& Innovation. BL is supported by the European Research Council (ERC-StG-716532-PUNCA) and the UK Science and Technology Facilities Council through grant ST/P000541/1.
\end{acknowledgments}

\newcommand{\aj}{Astron.\ J.}
\newcommand{\apjl}{\apj\ Lett.}
\newcommand{\jcap}{Journal of Cosmology and Astroparticle Physics}
\newcommand{\mnras}{MNRAS}


\bibliography{robust}

\end{document}